\documentclass[showpacs, prb, aps, superscriptaddress, twocolumn,10pt]{revtex4}
\usepackage{mathrsfs}
\usepackage{array}
\usepackage{amsmath}
\usepackage{graphicx}
\usepackage{amstext}
\usepackage{amsfonts}
\usepackage{bm}

\begin{document}
\title{Master equation approach to transient quantum transport incorporating with initial correlations}
\author{Pei-Yun Yang}
\affiliation{Department of Physics and Center for Quantum
information Science, National Cheng Kung University, Tainan 70101,
Taiwan}
\author{Chuan-Yu Lin}
\affiliation{Department of Physics and Center for Quantum
information Science, National Cheng Kung University, Tainan 70101,
Taiwan}
\author{Wei-Min Zhang}
\email{wzhang@mail.ncku.edu.tw}
\affiliation{Department of Physics
and Center for Quantum information Science, National Cheng Kung
University, Tainan 70101, Taiwan}
\begin{abstract}
In this paper, the exact transient quantum transport of
non-interacting nanostructures is investigated in the presence of
initial system-lead correlations and initial lead-lead correlations
for a device system coupled to general electronic leads. The exact
master equation incorporating with initial correlations is derived
through the extended quantum Langevin equation. The effects of the
initial correlations are manifested through the time-dependent
fluctuations contained explicitly in the exact master equation. The
transient transport current incorporating with initial correlations
is obtained from the exact master equation. The resulting transient
transport current can be expressed in terms of the single-particle
propagating and correlation Green functions of the device system. We
show that the initial correlations can affect quantum transport not
only in the transient regime, but also in the steady-state limit
when system-lead couplings are strong enough so that electron
localized bound states occur in the device system.
\end{abstract}

\maketitle
\section{Introduction}
Quantum transport incorporating with initial correlations in
nanostructures is a long-standing problem in mesoscopic
physics.\cite{Haug2008} In the past two decades, investigations of
quantum transport are mainly focused on steady-state
phenomena,\cite{Datta1995,Blanter2000,Imry2002} where initial
correlations are not essential due to the memory loss effect. Recent
experimental developments allow one to measure transient quantum
transport in different nano and quantum devices.\cite{Lu03422,
Bylander05361,Gustavsson08152101} In the transient transport regime,
initial correlations could induce different transport effects. In
this paper, using the exact master equation
approach.\cite{Tu2008,Jin2010} we shall attempt to address the
transient quantum transport incorporating with initial correlations.

Conventional approaches for studying quantum transport include the
scattering theory,\cite{Buttiker1992,Imry2002} the nonequilibrium
Green function technique,\cite{Wingreen1993,Haug2008} and the master
equation
approach.\cite{Schoeller1994,Jin2008a,Tu2008,Jin2010,Li2005} In the
steady-state quantum transport regime, the famous
Landauer-B\"{u}ttiker formula\cite{Buttiker1992} has been widely
utilized to calculate successfully various transport properties in
semiconductor nanostructures.\cite{Datta1995,Imry2002} In the
Landauer-B\"{u}ttiker formula, the transport current is given in
terms of a simple transmission coefficient obtained from the
single-particle scattering matrix. However, the scattering theory
considers the reservoirs connecting to the scattering region (the
device system) to be always in equilibrium and electrons in the
reservoir are always incoherent. Thus, the Landauer-B\"{u}ttiker
formula becomes invalid to transient quantum transport. The
scattering theory method could be extended to deal with
time-dependent transport phenomena, through the so-called the
Floquet scattering theory,\cite{Moskalets2004} but it is only
applicable to the case of the time-dependent quantum transport for
systems driven by periodic time-dependent external fields.

The nonequilibrium Green function technique based on Keldysh
formalism\cite{Schwinger1961} has also been used extensively to
investigate the steady-state quantum transport in mesoscopic
systems, where the initial correlations were also
ignored.\cite{Wingreen1993,Haug2008} In comparison with the
scattering theory approach, the nonequilibrium Green function
technique provides a more microscopic picture to electron transport
by formulating the transport current or more specifically the
transmission coefficients in terms of the nonequilibrium electron
Green functions of the device system in the
nanostructures.\cite{Haug2008} Wingreen {\it et al.}~extended
Keldysh's nonequilibrium Green function technique to time-dependent
quantum transport under time-dependent external bias and gate
voltages.\cite{Wingreen1993} But in the Keldysh formalism,
nonequilibrium Green functions are defined with the initial time
$t_0 \rightarrow -\infty$, where the initial correlations are hardly
taken into account, which makes the technique to be useful mostly in
the nonequilibrium steady-state regime.

The transient quantum transport was first proposed by Cini,
\cite{Cini805887} under the so-called partition-free scheme. In this
scheme, the whole system (the device system plus the leads together)
is in thermal equilibrium up to time $t=0$, and then one applies the
external bias to let electrons flow. Thus, the device system and the
leads are initially correlated. Stefanucci {\it et.
al.}~\cite{Stef04195318,Stef07195115} adapted nonequilibrium Green
functions with the Kadanoff-Baym formalism\cite{Kadanoff1962} to
investigate the transient quantum transport with the partition-free
scheme proposed by Cini.\cite{Cini805887} They obtained an analytic
transient transport current in the wide-band limit. But in these
works,\cite{Stef04195318,Stef07195115} the transport solution is
given with the nonequilibrium Green functions of the total system,
rather than the Green functions for the device part of the
nanostructure.\cite{Wingreen1993} Thus the advantage of simplifying
the microscopic picture of quantum transport in terms of
nonequilibrium Green functions of the device system may not be so
obvious as in the Meir-Wingreen formula.\cite{Meir1992}

On the other hand, the master equation
approach\cite{Schoeller1994,Gurvitz1996} concerns the dynamic
properties of the device system in terms of the time evolution of
the reduced density matrix $\rho(t)=\rm{Tr_E}\rho_{tot}(t)$, where
$\rm{Tr_E}$ is the trace over the environmental degrees of freedom.
The dissipation and fluctuation dynamics of the device system
induced by the reservoirs are fully manifested through the master
equation. Also, all the transport properties can be derived from the
master equation. In principle, the master equation for the quantum
transport can be investigated in terms of the real-time diagrammatic
expansion approach up to all the orders.\cite{Schoeller1994}
However, most of master equations are obtained by the perturbation
theory up to the second order of the system-lead couplings, which is
mainly applicable in the sequential tunneling regime.\cite{Li2005}
An interesting development of master equations in quantum transport
systems is the hierarchical expansion of the equations of motion for
the reduced density matrix,\cite{Jin2008a} which provided a
systematical and also quite useful numerical calculation scheme to
quantum transport.

A few years ago, we derived the exact master equation for
non-interacting nano-devices,\cite{Tu2008,Jin2010} using the
Feynman-Vernon influence functional approach\cite{FeynmanVernon} in
the coherent-state representation. The obtained exact master
equation not only describes the quantum state dynamics of the device
system but also takes into account all the transient electronic
transport properties. The transient transport current can be
directly obtained from the exact master equation, which turns out to
be expressed by the nonequilibrium Green functions of the device
system. This unifies the master equation approach and the
nonequilibrium Green function technique for quantum transport.
However, the exact master equation given in
Ref.~[\onlinecite{Tu2008,Jin2010}] is derived in the partitioned
scheme in which the system and the leads are initially uncorrelated.
The result transport current is consistent with the Meir-Wingreen
formula.\cite{Wingreen1993} This new theory has also been used to
study quantum transport (including the transient transport) for
various nanostructures recently.\cite{Tu2008,Tu2012} However,
realistically, it is possible and often unavoidable in experiments
that the device system and the leads are initially correlated.
Therefore, the transient transport theory based on master equation
that takes the effect of initial correlations into account deserves
of a further investigation.

In this paper, we obtain the exact master equation including the
effect of initial correlations for non-interacting nanostructures
through the extended quantum Langevin equation.\cite{Yang14115411}
We find that the initial correlations only affect on the fluctuation
dynamics of the device system, while the dissipation dynamics
remains the same as in the case of initially uncorrelated states.
The transient transport current in the presence of initial
system-lead and lead-lead correlations are also obtained directly
from the exact master equation. This transient transport theory is
suitable for arbitrary initial states between the system and its
reservoirs, so that it naturally covers both the partitioned and
partition-free schemes studied in previous
works.\cite{Wingreen1993,Stef04195318, Stef07195115} Taking an
experimentally realizable nano-fabrication system, a single-level
quantum dot coupled to two one-dimensional tight-binding leads, as
an specific example, we examine in detail the initial correlation
effects in the transient transport current as well as in the
electron occupation of the device system.

In particular, we find that the electron occupation and transport
current in the transient and also in the steady-state regimes can be
different for partitioned and partition-free schemes when the
system-lead couplings become strong, where localized bound states
will appear in the dot system. The quantum transport in the presence
of localized bound states has also been studied in the
literatures.\cite{boundstate1,boundstate2,boundstate3,
Dhar06085119,Stef07195115} In fact, Dhar and Sen considered a wire
connected to reservoirs that is modeled by a tight-binding
noninteracting Hamiltonian in the partitioned
scheme,\cite{Dhar06085119} and they gave the steady-state solution
of the density matrix and the current. Their results show that the
memory effects induced by the localized bound states can be observed
such that the density matrix of the system is initial-state
dependent. Stefanucci used the Kadanoff-Baym formalism to formally
study the localized bound state effects in the quantum transport in
the partition-free scheme.\cite{Stef07195115} He found that the
biased system with localized bound states does not evolve toward a
stationary state. In this paper, we present the detailed time
evolution of the density matrix and the transient transport current
for both the partitioned and partition-free schemes. The
initial-state dependance of the density matrix in the partitioned
scheme is obvious in our formalism, as given in Eq.~(\ref{rho11}).
Meanwhile, the result of the system being unable to reach a
stationary state found by Stefanucci\cite{Stef07195115} is also
reproduced in our calculation. Moreover, we find that this result is
sensitive to the applied bias, as we will show in the paper.

The rest of the paper is organized as following. In Sec. II, the
transient quantum transport with initial correlations is formulated
for non-interacting device system coupled to leads through the
master equation approach. In Sec. III, the transport dynamics with
or without the initial correlations are explored in detail with a
quantum dot device coupled to two leads modeled by one-dimensional
tight-binding chains. The initial correlation effects in the
electron occupation and in the transient transport current are
explicitly calculated. Finally, a summary and discussion are given
in Sec.IV.

\section{Master equation with initial correlations}
To study transient quantum transport incorporating with initial
correlations, we utilize the master equation approach. Consider a
nanostructure consisting of a quantum device coupled with two leads
(the source and drain), which can be described by a Fano-Anderson
Hamiltonian,
\begin{align}
H(t)=&H_S(t)+H_E(t)+H_{SE}(t)\notag\\
=&\sum_{ij}\bm{\varepsilon}_{ij}(t) a^\dag_i a_j+\sum_{\alpha
k}\epsilon_{\alpha k}(t) c_{\alpha k}^\dag c_{\alpha k}\notag\\
&+\sum_{i \alpha k} \big[V_{i \alpha k}(t)a^\dag_i c_{\alpha
k}+V^*_{i \alpha k}(t)c_{\alpha k}^\dag a_i \big] ,
\label{Hamiltonian}
\end{align}
where the electron-electron interactions are not considered. Here,
$a_{i}^{\dag}$ ($a_{i}$) and $c_{\alpha k}^{\dag}$ ($c_{\alpha k}$)
are creation (annihilation) operators of electrons in the device
system and the lead $\alpha$, respectively;
$\bm{\varepsilon}_{ij}(t)$ and $\epsilon_{\alpha k}(t)$ are the
corresponding energy levels, and $V_{i\alpha k}(t)$ is the tunneling
amplitude between the orbital state $i$ in the device system and the
orbital state $k$ in the lead $\alpha$. These time-dependent
parameters in Eq.~(\ref{Hamiltonian}) can be controlled by external
bias and gate voltages in experiments.

Because the system and the leads are coupled only by the electron
tunneling effect, and the leads are made by electrodes, the master
equation describing the time evolution of the reduced density matrix
$\rho(t)= \rm{Tr}_E[\rho_{tot}(t)]$ should have in general the
following form:\cite{Tu2008,Tu2012,Jin2010,Yang14115411}
\begin{align}
{d\rho(t)\over dt} =& -i\big[H'_S(t),\rho(t)\big]+\sum_{ij}\big\{
\bm{\gamma}_{ij}(t)\big[2a_j\rho(t) a_i^{\dag} \notag \\
& -a_i^{\dag}a_j\rho(t) -\rho(t) a_i^{\dag}a_j\big]
 +\widetilde{\bm{\gamma}}_{ij}(t)\big[ a_i^{\dag}\rho(t)a_j \notag \\
& -a_j\rho(t)a_i^{\dag}+a_i^{\dag}a_j\rho(t) -\rho(t)
a_ja_i^{\dag}\big] \big\}\notag\\
=&-i\big[H_S(t),\rho(t)\big]+\sum_{\alpha}\big[\mathcal{L}^{+}_{\alpha}(t)+\mathcal{L}^{-}_{\alpha}(t)\big]\rho(t),
\label{Master Equation}
\end{align}
where the renormalized Hamiltonian
$H'_S(t)=\sum_{ij}\bm{\varepsilon}'_{ij}(t)a_i^{\dag}a_j$, the
coefficient $\bm{\varepsilon}'_{ij}(t)$ is the corresponding
renormalized energy matrix of the device system, including the
energy shift of each level and the lead-induced couplings between
different levels. The time-dependent dissipation coefficients
$\bm{\gamma}_{ij}(t)$ and the fluctuation coefficients
$\widetilde{\bm{\gamma}}_{ij}(t)$ take into account all the
back-actions between the device system and the reservoirs. The
current superoperators of lead $\alpha$,
$\mathcal{L}^{+}_{\alpha}(t)$ and $\mathcal{L}^{-}_{\alpha}(t)$ give
the transport current from lead $\alpha$ to the device system such
that
\begin{align}
I_\alpha(t)=&-e\langle\frac{dN_\alpha(t)}{dt}\rangle\notag\\
=&e\rm{Tr}\big[\mathcal{L}^{+}_{\alpha}(t)\rho(t)\big]=\!-e\rm{Tr}\big[\mathcal{L}^{-}_{\alpha}(t)\rho(t)\big],
\end{align}
where $N_\alpha(t)=\sum_k c^\dag_{\alpha k}(t)c_{\alpha k}(t)$ is
the particle number of the lead $\alpha$.

\subsection{Without initial correlations}
When the whole system is initially partitioned, namely,
$\rho_{tot}(t_0)=\rho(t_0) \otimes \rho_E(t_0)$, in which the system
can be in arbitrary initial state $\rho(t_0)$ and the leads are
initially at equilibrium
$\rho_E(t_0)=\frac{1}{Z}e^{-\sum_\alpha\beta_\alpha(
H_\alpha-\mu_\alpha N_\alpha)}$. In this case, there is no
system-lead and lead-lead correlations in the initial state. Then
the time-dependent coefficients in Eq.~(\ref{Master Equation}) can
be exactly derived using the Feynman-Vernon influence functional
approach in fermion coherent state
representation,\cite{Jin2010,Tu2008} with the results
\begin{subequations}
\label{ecoff1}
\begin{align}
\bm{\varepsilon}'_{ij}(t)= &
\frac{i}{2}\big[\dot{\bm{u}}(t,t_0)\bm{u}^{-1}(t,t_0) - {\rm
H.c.}\big]_{ij}
\notag\\&=\bm{\varepsilon}_{ij}(t)-\frac{i}{2}\sum_\alpha[\bm{\kappa}_\alpha(t)-\bm{\kappa}^\dag_\alpha(t)]_{ij},\\
\bm{\gamma}_{ij}(t)= & -\frac{1}{2}\big[\dot{\bm{u}}(t,t_0)\bm{u}^{-1}(t,t_0) + {\rm H.c.}\big]_{ij}
\notag\\&=\frac{1}{2}\sum_\alpha[\bm{\kappa}_\alpha(t)+\bm{\kappa}^\dag_\alpha(t)]_{ij},\\
\widetilde{\bm{\gamma}}_{ij}(t)=&\, \, \frac{d}{dt}\bm{v}_{ij}(t,t)
-[\dot{\bm{u}}(t,t_0)\bm{u}^{-1}(t,t_0)\bm{v}(t,t)+ \rm
H.c.]_{ij}\notag\\&=-\sum_\alpha[\bm{\lambda}_\alpha(t)+\bm{\lambda}^\dag_\alpha(t)]_{ij},
\end{align}
\end{subequations}
The current superoperators of lead $\alpha$,
$\mathcal{L}^{+}_{\alpha}(t)$ and $\mathcal{L}^{-}_{\alpha}(t)$, are
explicitly given by
\begin{subequations}
\begin{align}
\mathcal{L}^{+}_{\alpha}(t)\rho(t)=&-\sum_{ij}\big\{\bm{\lambda}_{\alpha
ij}(t)\big[a^\dag_ia_j\rho(t)+a^\dag_i\rho(t)a_j\big]\notag\\&+\bm{\kappa}_{\alpha
ij}(t)a^\dag_ia_j\rho(t)+\rm H.c.\big\},\\
\mathcal{L}^{-}_{\alpha}(t)\rho(t)=&\sum_{ij}\big\{\bm{\lambda}_{\alpha
ij}(t)\big[a_j\rho(t)a^\dag_i+\rho(t)a_ja^\dag_i\big]\notag\\&+\bm{\kappa}_{\alpha
ij}(t)a_j\rho(t)a^\dag_i+\rm H.c.\big\}.
\end{align}
\end{subequations}
In Eq.~(\ref{ecoff1}), $\bm{u}(t,t_0)$ and $\bm{v}(\tau,t)$ are
related to the nonequilibrium Green functions of the device system
in the  nonequilibrium formalism.\cite{Kadanoff1962} These Green
functions obey the following Kadanoff-Baym integro-differential
equations,
\begin{subequations}
\label{greenfn}
\begin{align}
&\frac{d}{d\tau}\bm{u}(\tau,t_0) +
i\bm{\varepsilon}\bm{u}(\tau,t_0)+
\sum_{\alpha}\int_{t_0}^{\tau}d\tau'\bm{g}_{\alpha}(\tau,\tau')\bm{u}(\tau',t_0) = 0,\\
&\frac{d}{d\tau}\bm{v}(\tau,t) + i\bm{\varepsilon}\bm{v}(\tau,t) +
\sum_{\alpha}\int_{t_0}^{\tau}d\tau'\bm{g}_{\alpha}(\tau,\tau')\bm{v}(\tau',t) \notag \\
& ~~~~~~~~~~~~~~~~~~~~~~~~~~~~~~~=
\sum_{\alpha}\int_{t_0}^td\tau'\bm{\widetilde{g}}_{\alpha}(\tau,\tau')\bm{u}^{\dag}(t,\tau'),
\end{align}
\end{subequations}
subject to the boundary conditions $\bm{u}(t_0,t_0)=1$ and $
\bm{v}(t_0,t)=0$ with $t_0\leq\tau\leq t$. Here, the self-energy
correlation functions from the lead to the device system,
$\bm{g}_{\alpha}(\tau,\tau')$ and
$\bm{\widetilde{g}}_{\alpha}(\tau,\tau')$, are found to be
\begin{subequations}
\label{selfenergycorrelation}
\begin{align}
\bm{g}_{\alpha ij}(\tau,\tau') &= \sum_{k}V_{i\alpha
k}(\tau)V^*_{j\alpha
k}(\tau')e^{-i\int_{\tau'}^{\tau}\epsilon_{\alpha
k}(\tau_1)d\tau_1},\\
\widetilde{\bm{g}}_{\alpha ij}(\tau,\tau')&=\sum_{k}V_{i\alpha
k}(\tau)V^*_{j\alpha k}(\tau')f_\alpha(\epsilon_{\alpha
k})e^{-i\int_{\tau'}^{\tau}\epsilon_{\alpha k}(\tau_1)d\tau_1}.
\end{align}
\end{subequations}
In Eq.~(\ref{selfenergycorrelation}b), $f_{\alpha}(\epsilon_{\alpha
k})=[e^{\beta_{\alpha}(\epsilon_{\alpha k}-\mu_{\alpha})}+1]^{-1}$
is the Fermi-Dirac distribution of lead $\alpha$ at initial time
$t_0$. Solving the inhomogeneous equation Eq.~(\ref{greenfn}b) with
the initial condition $\bm{v}(t_0,t)=0$, we obtain
\begin{align}
\bm{v}(\tau,t) =\sum_\alpha\int_{t_0}^{\tau} \!\!\! d\tau' \!\!
\int_{t_0}^t \!\!\! d\tau''
\bm{u}(\tau,\tau')\widetilde{\bm{g}}_\alpha(\tau',\tau'')
\bm{u}^{\dag}(t,\tau''). \label{v}
\end{align}
In fact, $\bm{u}_{ij}(t,t_0)=\langle \{ a_i(t), a^\dag_j
(t_0)\}\rangle$ is the electron spectral Green function and $
\bm{v}(\tau,t)$ is the electron correlation Green function of the
device system in the non-equilibrium Green function
technique.\cite{Zhang12170402} The functions
$\bm{\kappa}_{\alpha}(t)$ and $\bm{\lambda}_{\alpha}(t)$ in
Eq.~(\ref{ecoff1}) can be solved from Eq.~(\ref{greenfn}),
\begin{subequations}
\label{kappalambda}
\begin{align}
\bm{\kappa}_\alpha(t)&=\int^t_{t_0}d\tau\bm{g}_\alpha(t,\tau)\bm{u}(\tau,t_0)[\bm{u}(t,t_0)]^{-1},
\\\bm{\lambda}_\alpha(t)=&\int^t_{t_0}d\tau[\bm{g}_\alpha(t,\tau) \bm{v}(\tau,t)-\widetilde{\bm{g}}_\alpha(t,\tau)\bm{u}^\dag(t,\tau)]\notag\\
&-\bm{\kappa}_\alpha(t)\bm{v}(t,t).
\end{align}
\end{subequations}
The transient transport current flowing from lead $\alpha$ to the
device system can be obtained directly from the current
superoperators in the master equation (\ref{Master Equation}), and
it is
\begin{align}
I_\alpha(t)
=&-e\rm{Tr}[\bm{\lambda}_\alpha(t)+\bm{\kappa}_\alpha(t)\bm{\rho}^{(1)}(t)+H.c.]\notag\\
=&-\!\! 2e{\rm Re}{\rm Tr}\!\!\int_{t_0}^t\!\!\! d\tau
[\bm{g}_\alpha(t,\tau)\bm{\rho}^{(1)}(\tau, t)
-\bm{\widetilde{g}}_\alpha(t,\tau)\bm{u}^{\dag}(t,\tau)] .\label{II}
\end{align}
In Eq.~(\ref{II}), the single-particle correlation function of the
device system $\bm{\rho}^{(1)}(\tau,t)$  is given by
\begin{align}
\bm{\rho}^{(1)}_{ij}(\tau,t) &=\langle
a^\dag_j(t)a_i(\tau)\rangle\notag\\
=&\big[\bm{u}(\tau,t_0)\bm{\rho}^{(1)}(t_0)\bm{u}^{\dag}(t,t_0)+
\bm{v}(\tau,t)\big]_{ij} , \label{rho11}
\end{align}
where $\bm{\rho}^{(1)}_{ij}(t_0)=\langle
a^\dag_j(t_0)a_i(t_0)\rangle$, is the initial single-particle
density matrix, and
$\bm{\rho}^{(1)}_{ij}(\tau,t)=-iG^{<}_{ij}(\tau,t)$ is indeed the
lesser Green function in the nonequilibrium Green function
technique, including a term induced by the initial occupation of the
device system (i.e. the first term in Eq.~(\ref{rho11})). Thus, the
transient transport current obtained from the master equation has
the exact same formula as that using the nonequilibrium Green
function technique,\cite{Wingreen1993} except for the above
mentioned term originated from the initial occupation of the device
system which was not considered in Ref.~[\onlinecite{Wingreen1993}].

\subsection{Including initial correlations}
When the device system and the leads are initially correlated, i.e.
$\rho_{tot}(t_0)\neq\rho(t_0) \otimes \rho_E(t_0)$, it would be
challenging to use the Feynman-Vernon influence functional approach
to derive the master equation. Alternately, in our previously
work,\cite{Yang14115411} we have used the extended quantum Langevin
equation to correctly determine the time-dependent coefficients in
the master equation in the absence of initial correlations. Here, we
shall use the same technique to determine the time-dependent
coefficients in the master equation when the initial correlations
are presented. The extended quantum Langevin equation for the device
system operators can be derived from the Heisenberg equation of
motion:\cite{Yang14115411}
\begin{align}
\frac{d}{dt}a_i(t)=&-i\sum_j \bm{\varepsilon}_{ij}(t) a_j(t)
-\sum_{\alpha j}\int_{t_0}^t d\tau \bm{g}_{\alpha
ij}(t,\tau)a_j(\tau) \notag\\& -i\sum_{\alpha k} V_{i\alpha k}(t)
c_{\alpha k}(t_0)e^{-i\int_{t_0}^t \bm{\epsilon}_{\alpha k}
(\tau_1)d\tau_1}, \label{lat}
\end{align}
where the first term gives the evolution of the device system
itself. The second term is the dissipation which arises from the
coupling between the device system and the leads, and which is
independent of initial state. As a result, the initial correlations
cannot affect on dissipation dynamics. The last term is the
fluctuation induced by the reservoirs (the leads), which depends
explicitly on the initial state of the whole system, so does initial
correlations. The non-local time-correlation function
$\bm{g}_{\alpha ij}(t,\tau)$ in Eq.~(\ref{lat}) is given in
Eq.~(\ref{selfenergycorrelation}a), which characterizes the
dissipation induced by the back-actions from the lead $\alpha$ to
the device system. Since this quantum Langevin equation is derived
exactly from Heisenberg equation of motion, it is valid for
arbitrary initial state of the device system and the leads,
including the case of the device system and the leads being
initially correlated.

Because of the linearity in the extended quantum Langevin equation
(\ref{lat}), its general solution can be expressed as
\begin{align}
a_i(t)=\sum_j \bm{U}_{ij}(t,t_0)a_j(t_0)+F_i(t) , \label{att}
\end{align}
where $\bm{U}_{ij}(t,t_0)$ is the single-particle electron
propagating function, and $F_i(t)$ is an color-noise force induced
by the electronic reservoirs (the leads). From Eq.~(\ref{lat}), one
can find that $\bm{U}_{ij}(t,t_0)$ and $F_i(t)$ must obey the
following equations,
\begin{subequations}
\label{equationofmotion}
\begin{align}
\frac{d}{dt}\bm{U}_{ij}(t,t_0) &+
i\big[\bm{\varepsilon}\bm{U}(t,t_0)\big]_{ij}\notag\\
&+\sum_{\alpha}\int_{t_0}^{t}d\tau\big[\bm{g}_{\alpha}(t,\tau)\bm{U}(\tau,t_0)\big]_{ij} = 0,\\
\frac{d}{dt}F_i(t)+i &\sum_{j}\bm{\varepsilon}_{ij}(t) F_j(t)
+\sum_{\alpha j}\int_{t_0}^t d\tau \bm{g}_{\alpha ij}(t,\tau)
F_j(\tau) \label{ft} \notag\\
&=-i \sum_{\alpha k}V_{i\alpha k}(t) c_{\alpha k}(t_0)
e^{-i\int_{t_0}^t\epsilon_{\alpha k}(\tau_1) d\tau_1},
\end{align}
\end{subequations}
subject to the initial conditions $\bm{U}_{ij}(t_0,t_0)=\delta_{ij}$
and $F_i(t_0)=0$. It is easy to prove that the propagating function
$\bm{U}_{ij}(t,t_0)$ is equal to the spectral Green function
$\bm{u}_{ij}(t,t_0)$ of Eq.~(\ref{greenfn}a):
\begin{align}
\bm{u}_{ij}(t,t_0)=\langle \{ a_i(t), a^\dag_j
(t_0)\}\rangle=\bm{U}_{ij}(t,t_0),
\end{align}
and the solution of Eq.~(\ref{equationofmotion}b) is given by
\begin{align}
F_i(t)\!\!=\!\!-i\! \sum_{\alpha k j}\! \int_{t_0}^t\!\!\!\! d\tau\!
\big[\bm{u}_{ij}(t,\tau)V_{j\alpha k}(\tau)
e^{-i\int_{t_0}^{\tau}\epsilon_{\alpha k}(\tau_1) d\tau_1}\big]\!
c_{\alpha k}(t_0). \label{Ft}
\end{align}

To determine the time-dependent coefficients in the master equation
(\ref{Master Equation}), we compute the equation of motion of the
single-particle density matrix of the device system,
$\bm{\rho}^{(1)}_{ij}(t)=\langle a_j^\dag(t) a_i(t) \rangle = {\rm
Tr}[a_j^\dag a_i \rho(t)]$. From the master equation (\ref{Master
Equation}), the result is given by
\begin{align}
\frac{d}{dt}\bm{\rho}^{(1)}_{ij}(t)=&\big\{\bm{\rho}^{(1)}(t)
\big[i\bm{\varepsilon}'(t)-\bm{\gamma}(t) \big]\big\}_{ij}
\notag\\&-\big\{\big[i\bm{\varepsilon}'(t) + \bm{\gamma}(t) \big]
\bm{\rho}^{(1)}(t)\big\}_{ij} + \widetilde{\bm{\gamma}}_{ij}(t) .
\label{aa1}
\end{align}
On the other hand, this equation can also be derived from the
quantum Langevin equation. Explicitly, the single-particle
correlation function of the device system calculated from the
solution Eq.~(\ref{att}) is given by
\begin{align}
\bm{\rho}^{(1)}_{ij}(\tau,t)&= \langle a^\dag_j(t) a_i(\tau)\rangle
\notag\\&
=\big[\bm{u}(\tau,t_0)\bm{\rho}^{(1)}(t_0)\bm{u}^{\dag}(t,t_0)\big]_{ij}
+\bm{v}'_{ij}(\tau,t) .\label{rho1}
\end{align}
Here,
\begin{align}
\bm{v}'_{ij}(\tau, t)&=\bm{\nu}_{ij}(\tau,t)\!+\!
\big[\bm{\chi}(\tau,t_0)
\bm{u}^{\dag}(t,t_0)+\bm{u}(\tau,t_0)\bm{\chi}^{\dag}(t,t_0)\big]_{ij},
\end{align}
with
\begin{subequations}
\label{a}
\begin{align}
\bm{\chi}_{ij}(t,t_0)=&\langle a_j^\dag(t_0)F_i(t)\rangle =-\!\sum_{\alpha}\!\int_{t_0}^t d\tau [\bm{u}(t,\tau)\bm{g}^\chi_\alpha(\tau,t_0)]_{ij},\\
\bm{\nu}_{ij}(\tau, t)=&\langle
F_j^\dag(t)F_i(\tau)\rangle \notag \\
=&\!\!\sum_{\alpha}\!\! \int_{t_0}^{\tau} \!\!d\tau'
\!\!\!\int_{t_0}^{t} d\tau'' [\bm{u}(\tau, \tau')
\widetilde{\bm{g}}'_{\alpha}(\tau', \tau'') \bm{u}^{\dag}(t,\tau'')]_{ij} ,\\
\bm{g}^\chi_{\alpha ij}(\tau,t_0)=&i\sum_{k} V_{i\alpha k}(\tau)
e^{-i\int_{t_0}^{\tau} \epsilon_{\alpha k}(\tau_1)d\tau_1}\langle
a_{j}^\dag(t_0)c_{\alpha k}(t_0)\rangle,\\
\widetilde{\bm{g}}'_{\alpha ij}(\tau
,\tau')=&\sum_{\alpha'}\sum_{kk'} V_{i\alpha
k}(\tau)e^{-i\int_{t_0}^{\tau}\epsilon_{\alpha
k}(\tau_1)d\tau_1} \notag \\
&\times\! V^*_{j\alpha'
k'}(\tau')e^{i\int_{t_0}^{\tau'}\epsilon_{\alpha'
k'}(\tau_1)d\tau_1}\langle c_{\alpha' k'}^\dag(t_0)c_{\alpha
k}(t_0)\rangle.
\end{align}
\end{subequations}
Physically, $\bm{g}^\chi_{\alpha}(\tau ,t_0)$ characterizes the
initial system-lead correlations, and $\widetilde{\bm{g}}'_{\alpha
}(\tau ,\tau')$ is a time-correlation function associated with the
initial electron correlations in the leads. Compare
Eq.~(\ref{rho11}) with Eq.~(\ref{rho1}), as one can see, the initial
correlations modify the last term $\bm{v}(\tau,t)$ in
Eq.~(\ref{rho11}) by $\bm{v}'(\tau, t)$. Eq.~(\ref{rho1}) indeed
gives the exact solution of the lesser Green function incorporating
with initial correlations.

From Eq.~(\ref{rho1}), we can find
\begin{align}
\frac{d}{dt} \bm{\rho}_{ij}^{(1)}(t)=&
[\dot{\bm{u}}(t,t_0)\bm{u}^{-1}(t,t_0)\bm{\rho}^{(1)}(t)+{\rm H.c.}]_{ij}\notag\\
&-[\dot{\bm{u}}(t,t_0)\bm{u}^{-1}(t,t_0)\bm{v}'(t,t)+{\rm H.c.}]_{ij}\notag\\
&+\frac{d}{dt}\bm{v}'_{ij}(t,t)\label{aa2} .
\end{align}
Now, by comparing Eq.~(\ref{aa1}) with Eq.~(\ref{aa2}), we can
uniquely determine the time-dependent renormalized energy
$\bm{\varepsilon}'_{ij}(t)$, the dissipation, and fluctuation
coefficients $\bm{\gamma}_{ij}(t)$ and
$\widetilde{\bm{\gamma}}_{ij}(t)$ in the master equation
incorporating with initial correlations. The results are given as
follows,
\begin{subequations}
\label{ecoff}
\begin{align}
\bm{\varepsilon}'_{ij}(t)&=
\frac{i}{2}\big[\dot{\bm{u}}(t,t_0)\bm{u}^{-1}(t,t_0) - {\rm
H.c.}\big]_{ij}\notag\\&=\bm{\varepsilon}_{ij}(t)-\frac{i}{2}\sum_\alpha[\bm{\kappa}_\alpha(t)-\bm{\kappa}^\dag_\alpha(t)]_{ij},\\
\bm{\gamma}_{ij}(t)&=
-\frac{1}{2}\big[\dot{\bm{u}}(t,t_0)\bm{u}^{-1}(t,t_0) + {\rm
H.c.}\big]_{ij}\notag\\&=\frac{1}{2}\sum_\alpha[\bm{\kappa}_\alpha(t)+\bm{\kappa}^\dag_\alpha(t)]_{ij},\\
\widetilde{\bm{\gamma}}_{ij}(t)&= \frac{d}{dt}\bm{v}'_{ij}(t,t)
-[\dot{\bm{u}}(t,t_0)\bm{u}^{-1}(t,t_0)\bm{v}'(t,t)+ \rm H.c.]_{ij}
\notag\\&=-\sum_\alpha[\bm{\lambda}'_\alpha(t)+\bm{\lambda}'^\dag_\alpha(t)]_{ij}.
\end{align}
\end{subequations}
From the above results, one can see that the renormalized energy and
the dissipation coefficients are independent of the initial
correlations and are identical to the results given in
Eq.~(\ref{ecoff1}a) and (\ref{ecoff1}b) for the decoupled initial
state. The fluctuation coefficients also have the same form as that
in the uncorrelated case, but the function $\bm{v}(t,t_0)$ which
characterizes thermal fluctuation of the leads in the uncorrelated
case is now modified by $\bm{v}'(t,t_0)$ which involves both the
initial system-lead and the initial lead-lead correlations. In other
words, initial correlations only contribute to the fluctuations
dynamics of the device system.

Correspondingly, $\bm{\kappa}_\alpha(t)$ in Eq.~(\ref{ecoff}) is the
same as Eq.~(\ref{kappalambda}a), but $\bm{\lambda}_\alpha(t)$ is
modified by the initial correlations:
\begin{align}
\bm{\lambda}'_\alpha(t)=&\int^t_{t_0}d\tau[\bm{g}_\alpha(t,\tau)\bm{v}'(\tau,t)-\widetilde{\bm{g}}'_\alpha(t,\tau)\bm{u}^\dag(t,\tau)]\notag\\
&-\bm{\kappa}_\alpha(t)\bm{v}'(t,t)+\bm{g}^\chi_\alpha(t,t_0)\bm{u}^\dag(t,t_0)
\end{align}
Thus, the transient transport current $I_\alpha(t)$ incorporating
with the initial correlations is given by
\begin{align}
I_\alpha(t)
=&-e\rm{Tr}[\bm{\lambda}'_\alpha(t)+\bm{\kappa}_\alpha(t)\bm{\rho}^{(1)}(t)+H.c.]\notag\\
=&- 2e{\rm Re}{\rm
Tr}\big[\int_{t_0}^t d\tau [\bm{g}_\alpha(t,\tau)\bm{\rho}^{(1)}(\tau, t) \notag\\
&-\widetilde{\bm{g}}'_\alpha(t,\tau)\bm{u}^{\dag}(t,\tau)]
+\bm{g}^\chi_\alpha(t,t_0) \bm{u}^{\dag}(t,t_0) \big].  \label{I}
\end{align}
where
$\widetilde{\bm{g}}'_{\alpha}(t,\tau)=\sum_{\alpha'}\widetilde{\bm{g}}'_{\alpha
\alpha'}(t,\tau)$. Eq.~(\ref{I}) contains all the information of
transient transport physics incorporating with initial correlations,
and is purely expressed in terms of nonequilibrium Green functions
of the device system. It is easy to see that, if the whole system is
in the partitioned scheme, then, $\bm{g}^\chi_\alpha(t,t_0)=0$ and
$\widetilde{\bm{g}}'_\alpha(t,\tau)$ is reduced to
$\widetilde{\bm{g}}_\alpha(t,\tau)$ in
Eq.~(\ref{selfenergycorrelation}b). The single-particle correlation
function Eq.~(\ref{rho1}) goes back to Eq.~(\ref{rho11}), and
Eq.~(\ref{I}) recovers the transient transport current
Eq.~(\ref{II}) for the partitioned scheme.

In conclusion, the master equation (\ref{Master Equation}) with the
time-dependent coefficients of Eq.~(\ref{ecoff}) can be used to
describe the non-Markovian dynamics of nano-device systems coupled
to leads involved initial system-lead and lead-lead correlations. It
should be pointed out that if the leads are made by superconductors
where there may be initial pairing correlations. Then, the form of
Eq.~(\ref{Master Equation}) may need to be modified further. We
leave this problem for further investigation. Nevertheless, the
master equation (\ref{Master Equation}) is sufficient for the
description of the transient quantum transport of nanostructures
with initial correlations. The memory effects, including the
initial-state dependence, are fully embedded into these
time-dependent dissipation and fluctuation coefficients in our
master equation.

\subsection{Relation to the nonequilibrium Green function of the total system}
In this subsection, as a consistent check, we prove that
Eq.~(\ref{rho1}) can be reexpressed to the result obtained by
Stefanucci and Almbladh in terms of the nonequilibrium Green
functions of the total system (i.e. Eq.(4) in
Ref.[\onlinecite{Stef04195318}]). As we have shown,
$\bm{u}_{ij}(t,t_0)$ is the spectral Green function,
$\bm{u}_{ij}(t,t_0)=\langle \{ a_i(t), a^\dag_j (t_0)\}\rangle$, the
retarded and advanced Green functions of the device system can be
expressed by $\bm{u}_{ij}(t,t_0)$:
\begin{align}
G^R_{ij}(t,t_0)&=-i\theta(t-t_0)\langle\{a_i(t),a^\dag_j(t_0)\}\rangle=-i\bm{u}_{ij}(t,t_0),\\
G^A_{ij}(t_0,t')&=i\theta(t'-t_0)\langle\{a_i(t_0),a^\dag_j(t')\}\rangle=i\bm{u}^\dag_{ij}(t',t_0).
\end{align}
Here, we have used the fact that the step function
$\theta(t-t_0)=\theta(t'-t_0)=1$ because $t$ and $t'$ are always
larger than initial time $t_0$. The initial single-particle density
matrix is associated with the lesser Green function of the device
system: $\bm{\rho}^{(1)}_{ij}(t_0)=-iG^<_{ij}(t_0,t_0)=\langle
a^\dag_j(t_0)a_i(t_0)\rangle$. On the other hand, the nonequilibrium
system-lead Green functions can be expressed by
$\bm{u}_{ij}(\tau,\tau')$:
\begin{subequations}
\begin{align}
G^R_{i,\alpha k}(t,t_0)&=-i\theta(t-t_0)\langle \{a_i(t),
c^\dag_{\alpha k}(t_0) \} \rangle \notag\\&=\! -\!\sum_l\!\!
\int^t_{t_0}\!\! d\tau' \bm{u}_{il}(t,\tau') V_{l\alpha
k}(\tau')e^{-i\int^{\tau'}_{t_0}\epsilon_{\alpha k}(\tau_1)d\tau_1},
\\G^A_{\alpha k, j}(t_0,t')&=i\theta(t'-t_0)\langle \{c_{\alpha k}(t_0),
a^\dag_j(t') \} \rangle \notag\\&=\! -\!\sum_l\!\!
\int^{t'}_{t_0}\!\! d\tau' e^{i\int^{\tau'}_{t_0}\epsilon_{\alpha
k}(\tau_1)d\tau_1} V^*_{l\alpha k}(\tau')\bm{u}^\dag_{lj}(t,\tau').
\end{align}
\end{subequations}

Thus, it is straightforward to find that
\begin{widetext}
\begin{align}
i\bm{\rho}^{(1)}_{ij}(t,t')=&i[\bm{u}(t,t_0)\bm{\rho}^{(1)}(t_0)\bm{u}^\dag(t',t_0)]_{ij}
+i[\bm{u}(t,t_0)\bm{\chi}^\dag(t',t_0)]_{ij}+i[\bm{\chi}(t,t_0)\bm{u}^\dag(t',t_0)]_{ij}+i\bm{\nu}_{ij}(t,t')\notag\\=&\sum_{i'j'}
G^R_{ii'}(t,t_0)G^<_{i'j'}(t_0,t_0)G^A_{j'j}(t_0,t')+\sum_{i'}\sum_{\alpha
k}G^R_{ii'}(t,t_0)G^<_{i',\alpha k}(t_0,t_0)G^A_{\alpha
k,j}(t_0,t')\notag\\&+\!\!\sum_{i'}\sum_{\alpha k}G^R_{i,\alpha
k}(t,t_0)G^<_{\alpha k,i'}(t_0,t_0)G^A_{i'j}(t_0,t')+\sum_{\alpha
k}\sum_{\alpha' k'}G^R_{i,\alpha k}(t,t_0)G^<_{\alpha
k,\alpha'k'}(t_0,t_0)G^A_{\alpha'k',j}(t_0,t')\notag\\=&[G^R(t,t_0)G^<(t_0,t_0)G^A(t_0,t')]_{ij}=G^<_{ij}(t,t'),
\end{align}
\end{widetext}
with
\begin{align}
G^<_{i',\alpha k}(t_0,t_0)&=i\langle c^\dag_{\alpha
k}(t_0)a_{i'}(t_0) \rangle, \notag\\G^<_{\alpha
k,i'}(t_0,t_0)&=i\langle a^\dag_{i'}(t_0)c_{\alpha k}(t_0) \rangle,
\end{align}
and
\begin{align}
G^<_{\alpha k,\alpha' k'}(t_0,t_0)=i\langle c^\dag_{\alpha'
k'}(t_0)c_{\alpha k}(t_0) \rangle.
\end{align}
This shows that the single-electron correlation function obtained by
the extended quantum Langevin equation gives the same results as the
lesser Green function obtained by Stefanucci in terms of the
nonequilibrium Green function of the total
system.\cite{Stef04195318}

\section{Illustration of initial correlation effects}
In this section, we utilize the theory developed in the previous
section to study the effect of initial correlations on transport
dynamics. To be specific, we consider an experimentally realizable
nano-fabrication system,  a single-level quantum dot coupled to the
source and the drain which are modeled by two one-dimensional
tight-binding leads. The Hamiltonian of the whole system is given by
\begin{align}
H(t)=&\, \, \varepsilon_ca^\dag a -\sum_\alpha (\lambda_{\alpha 1}
a^{\dag}c_{\alpha 1}+ \lambda^*_{\alpha 1} c_{\alpha
1}^{\dag}a)\notag\\&+\sum_{\alpha}\sum_{n=1}^{\mathcal{N}}
[\epsilon_{\alpha }+U_\alpha(t)] c_{\alpha n}^\dag c_{\alpha n}
\nonumber\\& -\sum_{\alpha}\sum_{n=1}^{\mathcal{N}-1}
(\lambda_{\alpha}c_{\alpha n}^{\dag}c_{\alpha
n+1}+\lambda^{*}_{\alpha}c_{\alpha n+1}^{\dag}c_{\alpha
n}),\label{H2}
\end{align}
where $a$ ($a^\dag$) is the annihilation (creation) operator of the
single-level dot with the energy level $\varepsilon_c$, and
$c_{\alpha n}$ ($c_{\alpha n}^\dag$) the annihilation (creation)
operator of lead $\alpha$ at site $n$. All the sites in lead
$\alpha$ have an equal on-site energy $\epsilon_{\alpha}$.
$U_\alpha(t)$ is the time-dependent bias voltage applied on lead
$\alpha$ to shift the on-site energy. The second term in
Eq.~(\ref{H2}) describes the coupling between the quantum dot and
the first site of the lead $\alpha$ with the coupling strength
$\lambda_{\alpha 1}$. The last term characterizes the electron
tunneling between two consecutive sites in the lead $\alpha$ with
tunneling amplitude $\lambda_{\alpha}$, and $\mathcal{N}$ is the
total number of sites.

Using the Fourier transformation, the electron operator at site $n$
can be expressed in the k-space,
\begin{eqnarray}
c_{\alpha n}=\sqrt{\frac{2}{\mathcal{N}}}e^{-in\phi}\sum_{k}\sin(n
k)c_{\alpha k},\label{frr}
\end{eqnarray}
where $k=\frac{\pi l}{\mathcal{N}}$,
$l=1\cdot\cdot\cdot\mathcal{N}$, corresponding to the Bloch modes,
and $\phi=\arg(\lambda_\alpha)$. Then, the Hamiltonian (\ref{H2})
becomes
\begin{align}
H(t)=&\varepsilon_ca^\dag a+\sum_{\alpha k}\epsilon_{\alpha
k}(t)c_{\alpha k}^\dag c_{\alpha k}\notag
\\&+\sum_{\alpha k}[V_{\alpha k}a^{\dag}c_{\alpha
k}+V^{*}_{\alpha k}c^{\dag}_{\alpha k}a], \label{H3}
\end{align}
where
\begin{subequations}
\begin{align}
\epsilon_{\alpha k}(t)&=\epsilon_{\alpha k}+U_\alpha(t)~,
~~\epsilon_{\alpha k}=\epsilon_\alpha-2|\lambda_\alpha|\cos k, \\
V_{\alpha k}&=-\sqrt{\frac{2}{\mathcal{N}}}\lambda_{\alpha
1}e^{-i\phi}\sin k. \label{strength}
\end{align}
\end{subequations}
The Hamiltonian (\ref{H3}) has the same structure as the Hamiltonian
(\ref{Hamiltonian}) in Sec.~II. In the following, we consider two
different initial states that are discussed the most in the
literatures. One is partition-free scheme in which the whole system
is in equilibrium before the external bias is switched on. The other
one, which does not have any initial correlation, is the partitioned
scheme in which the initial state of the dot system is uncorrelated
to the leads before the tunneling couplings are turned on. The dot
can be in any arbitrary initial state $\rho(t_0)$ and the leads are
initially at separated equilibrium state. Both of these two schemes
can be realized through different experimental setup. By comparing
the transient transport dynamics for these two different initial
schemes, one can see under what circumstances the initial
correlations will affect on quantum transport in the transient
regime as well as in the steady-state limit.

\subsection{Partition-free scheme}\label{case1}
In the partition-free scheme, the whole system is in equilibrium
before the external bias voltage $U_\alpha(t)$ is turned on. The
applied bias voltage is set to be uniform on each lead such that
$U_\alpha(t)=U_\alpha\Theta(t-t_0)$, so $H(t\leq t_0)=H$ is
time-independent. The density matrix of the whole system is given by
\begin{align}
\rho_{tot}(t_0)=\frac{1}{Z}e^{-\beta(H-\mu N)}
\end{align}
where $H$ and $N$ are respectively the total Hamiltonian and the
total particle number operator at initial time $t_0$. The whole
system is initially at the temperature $\beta=1/k_BT$ with the
chemical potential $\mu$. When $t>t_0$, we apply a uniform bias
voltage on each lead. The whole system then suddenly change into a
non-equilibrium state. When we take the site number $\mathcal{N}$ to
be infinity, the non-local time correlation function
$g_\alpha(t,\tau)$ can be expressed as
\begin{align}
g_\alpha(t,\tau)=\int\frac{d\epsilon}{2\pi}\Gamma_{\alpha}(\epsilon)e^{-i(\epsilon+U_\alpha)(t-\tau)},\label{g}
\end{align}
Here the spectral density can be explicitly calculated from the
tight-binding model with the result:
\begin{align}
\Gamma_\alpha(\epsilon) = \left\{
\begin{array}{ll}
\eta^2_\alpha\sqrt{4|\lambda_\alpha|^2-(\epsilon-\epsilon_\alpha)^2} & \mbox{if } |\epsilon-\epsilon_\alpha|\leq2|\lambda_\alpha|, \\
\\
0 & \mbox{otherwise,}
\end{array} \right. \label{gamma}
\end{align}
and $\eta_\alpha$ is the coupling ratio $|\lambda_{\alpha
1}|/|\lambda_\alpha|$ of lead $\alpha$.

To calculate the initial system-lead and lead-lead correlations,
$\langle a^\dag(t_0)c_{\alpha k}(t_0)\rangle$ and $\langle
c^\dag_{\alpha'k'}(t_0)c_{\alpha k}(t_0)\rangle$, we diagonalize the
total Hamiltonian of Eq.~(\ref{H3}) without the external bias
through the following transformation
\begin{subequations}
\label{eigenoperator}
\begin{align}
a(t_0)&=\sum_{\alpha k}V_{\alpha k}G^{R}(\epsilon_{\alpha
k})b_{\alpha k}, \\
c_{\alpha k}(t_0)&=b_{\alpha
k}+\sum_{\alpha'k'}G^{R}(\epsilon_{\alpha'k'})\frac{V_{\alpha'k'}V^*_{\alpha
k}}{\epsilon_{\alpha'k'}-\epsilon_{\alpha k}+i\delta}b_{\alpha'k'}.
\end{align}
\end{subequations}
Thus, without applying external bias  Eq.~(\ref{H3}) can be written
as
\begin{align}
H(t_0)=\sum_{\alpha k}\epsilon_{\alpha k}b^\dag_{\alpha k}b_{\alpha
k}.
\end{align}
In Eq.~(\ref{eigenoperator}), $\delta\rightarrow0^+$ and
$G^R(\epsilon_{\alpha k})$ is the retarded Green function of the
device system in the energy domain. The retarded (advanced) Green
function is defined as
\begin{align}
G^{R,A}(\epsilon_{\alpha k})=\frac{1}{\epsilon_{\alpha
k}-\varepsilon_c-\Sigma^{R,A}(\epsilon_{\alpha k})},
\end{align}
with the retarded (advanced) self-energy,
\begin{align}
\Sigma^{R,A}(\epsilon_{\alpha
k})=\sum_{\alpha'k'}\frac{|V_{\alpha'k'}|^2}{\epsilon_{\alpha
k}-\epsilon_{\alpha'k'}\pm i\delta}.
\end{align}
The initial system-lead and lead-lead correlations in the
partition-free scheme are then given by
\begin{widetext}
\begin{subequations}
\begin{align}
\langle a^\dag(t_0)c_{\alpha k}(t_0)\rangle=&V^*_{\alpha
k}G^A(\epsilon_{\alpha k})f(\epsilon_{\alpha k})
+\sum_{\alpha'k'}\frac{V^*_{\alpha
k}}{\epsilon_{\alpha'k'}-\epsilon_{\alpha
k}+i\delta}|V_{\alpha'k'}|^2G^R(\epsilon_{\alpha'k'})G^A(\epsilon_{\alpha'k'})f(\epsilon_{\alpha'k'}),\\
\langle c^\dag_{\alpha'k'}(t_0)c_{\alpha
k}(t_0)\rangle=&\delta_{\alpha
k,\alpha'k'}f(\epsilon_{\alpha k})\notag\\
&+G^R(\epsilon_{\alpha'k'})f(\epsilon_{\alpha'k'})\frac{V_{\alpha'k'}V^*_{\alpha
k}}{\epsilon_{\alpha'k'}-\epsilon_{\alpha k}+i\delta}
+G^A(\epsilon_{\alpha k})f(\epsilon_{\alpha
k})\frac{V_{\alpha'k'}V^*_{\alpha k}}{\epsilon_{\alpha
k}-\epsilon_{\alpha'k'}+i\delta}\notag\\
&+\sum_{\alpha_1k_1}|V_{\alpha_1k_1}|^2G^R(\epsilon_{\alpha_1k_1})G^A(\epsilon_{\alpha_1k_1})f(\epsilon_{\alpha_1k1})\frac{V_{\alpha'k'}}{\epsilon_{\alpha_1k_1}-\epsilon_{\alpha'k'}+i\delta}\frac{V^*_{\alpha
k}}{\epsilon_{\alpha_1k_1}-\epsilon_{\alpha'k'}+i\delta}.
\end{align}
\end{subequations}
\end{widetext}
Using the initial correlations obtained above, we can calculate the
initial density matrix in the dot, as well as the non-local time
system-lead and lead-lead correlation functions,
\begin{widetext}
\begin{subequations}
\label{initialcorrelations}
\begin{align}
\rho^{(1)}(t_0)=&\langle a^\dag(t_0)a(t_0)\rangle=\sum_{\alpha
k}|V_{\alpha k}|^2G^R(\epsilon_{\alpha k})G^A(\epsilon_{\alpha
k})f(\epsilon_{\alpha k})=\int\frac{d\epsilon}{2\pi}A(\epsilon)f(\epsilon),\\
g^\chi_\alpha(\tau,t_0)=&i\int\frac{d\epsilon}{2\pi}[\rm{Re}G^R(\epsilon)]
\Gamma_\alpha(\epsilon)f(\epsilon)e^{-i(\epsilon+U_\alpha)(\tau-t_0)}
+i\int\frac{d\epsilon}{2\pi}A(\epsilon)f(\epsilon)\big[\rm{P}\!\!\int\frac{d\epsilon'}{2\pi}\frac{\Gamma_\alpha(\epsilon')}{\epsilon-\epsilon'}e^{-i(\epsilon'+U_\alpha)(\tau-t_0)}\big],\\
g^v_\alpha(t,\tau)=&\int\frac{d\epsilon}{2\pi}\Gamma_\alpha(\epsilon)f(\epsilon)e^{-i(\epsilon+U_\alpha)(t-\tau)}\notag\\
&+\sum_{\alpha'}\int\frac{d\epsilon}{2\pi}A(\epsilon)f(\epsilon)\big[\mathcal{P}\int\frac{d\epsilon_1}{2\pi}\frac{\Gamma_{\alpha}(\epsilon_1)}{\epsilon-\epsilon_1}e^{-i(\epsilon_1+U_\alpha)(t-t_0)}\big]
\big[\mathcal{P}\int\frac{d\epsilon_2}{2\pi}\frac{\Gamma_{\alpha'}(\epsilon_2)}{\epsilon-\epsilon_2}e^{i(\epsilon_2+U_{\alpha'})(\tau-t_0)}\big]\notag\\
&+\sum_{\alpha'}\int\frac{d\epsilon}{2\pi}[\rm{Re}G^R(\epsilon)]\Gamma_{\alpha'}(\epsilon)f(\epsilon)e^{i(\epsilon+U_{\alpha'})(\tau-t_0)}\big[\mathcal{P}\int\frac{d\epsilon'}{2\pi}\frac{\Gamma_{\alpha}(\epsilon')}{\epsilon-\epsilon'}e^{-i(\epsilon'+U_{\alpha})(t-t_0)}\big]\notag\\
&+\sum_{\alpha'}\int\frac{d\epsilon}{2\pi}[\rm{Re}G^R(\epsilon)]\Gamma_{\alpha}(\epsilon)f(\epsilon)e^{-i(\epsilon+U_{\alpha})(t-t_0)}\big[\mathcal{P}\int\frac{d\epsilon'}{2\pi}\frac{\Gamma_{\alpha'}(\epsilon')}{\epsilon-\epsilon'}e^{i(\epsilon'+U_{\alpha'})(\tau-t_0)}\big]\notag\\
&-\frac{1}{4}\sum_{\alpha'}\int\frac{d\epsilon}{2\pi}A(\epsilon)\Gamma_{\alpha}(\epsilon)\Gamma_{\alpha'}(\epsilon)f(\epsilon)e^{-i(\epsilon+U_{\alpha})(t-t_0)}e^{i(\epsilon+U_{\alpha'})(\tau-t_0)},
\end{align}
\end{subequations}
\end{widetext}
where $\mathcal{P}$ represents the principal-value integral, and
$f(\epsilon)=1/(e^{\beta(\epsilon-\mu)}+1)$ is the Fermi
distribution of the total system after diagonalized the total
Hamiltonian. Thus, $f(\epsilon)$ involves all the initial
correlations in the partition-free scheme.

To simplify the calculation, we made the band center of the source
and drain match each other ($\epsilon_L=\epsilon_R=\epsilon_0$) and
take $\lambda_L=\lambda_R=\lambda_0$ to be the unit of energy. Also,
it should be pointed out that the diagonalization of
Eq.~(\ref{eigenoperator}) is performed in the absence of localized
states of the system. The localized bound states can be added back
via the spectral function
$A(\epsilon)=-2\rm{Im}G^R(\epsilon)$.\cite{Mahan1990} Without
applying the external bias, the spectral function $A(\epsilon)$ is
given by
\begin{align} \label{spectral}
A(\epsilon)=&2\pi\sum_{j=\pm}Z_j\delta(\epsilon-\epsilon_j)\Theta(\eta^2-\eta^2_\pm)\notag\\
\!\!\!-&\frac{2\rm{Im}\Sigma^{ret}(\epsilon)\Theta(4|\lambda_0|^2-(\epsilon-\epsilon_0)^2)}{[\epsilon-\epsilon_c-\rm{Re}\Sigma^{ret}(\epsilon)]^2+[\rm{Im}\Sigma^{ret}(\epsilon)]^2},
\end{align}
where $\eta^2=\eta^2_L+\eta^2_R$.  In Eq.~(\ref{spectral}), the
first term characterizes the localized bound state with energy
$\epsilon_j$ lying outside the energy band when the total coupling
ratio $\eta^2\geq\eta^2_\pm$, where
$\eta^2_\pm=2\mp\frac{\Delta}{|\lambda_0|}$ is the critical coupling
ratio, and $\Delta=\varepsilon_c-\epsilon_0$. As long as the energy
bands of the two leads overlap, there are at most two localized
bound states. The amplitude and the frequency of the localized bound
state are given by
\begin{subequations}
\begin{align}
Z_\pm&=\frac{1}{2}\frac{(\eta^2-2)\sqrt{4(\eta^2-1)|\lambda_0|^2+\Delta^2}\pm\eta^2\Delta}{(\eta^2-1)\sqrt{4(\eta^2-1)|\lambda_0|^2+\Delta^2}},\\
\epsilon_\pm&=\epsilon_0+\frac{(\eta^2-2)\Delta}{2(\eta^2-1)}\pm\frac{\eta^2\sqrt{4(\eta^2-1)|\lambda_0|^2+\Delta^2}}{2(\eta^2-1)},
\end{align}
\end{subequations}
The effect of localized bound states are manifested in the second
term of the system-lead and lead-lead correlation functions
Eq.~(\ref{initialcorrelations}b) and (\ref{initialcorrelations}b),
respectively. Thus, the effect of initial correlations will be
maintained in the steady-state limit. The second term in
Eq.~(\ref{spectral}) is the contribution from the regime of the
continuous band, which causes electron dissipation in the dot
system. The retarded self-energy in Eq.~(\ref{spectral}) is
\begin{align}
\Sigma^{\rm{ret}}(\epsilon)=\frac{1}{2}\eta^2[(\epsilon-\epsilon_0)-i\sqrt{4|\lambda_0|^2-(\epsilon-\epsilon_0)^2}].
\end{align}
The above calculations specify all quantities related to initial
correlations.

\subsection{Partitioned scheme}\label{case2}
For the partitioned scheme, the dot and the leads are initially
uncorrelated, and the leads are initially at equilibrium state
$\rho_E(t_0)=\frac{1}{Z}e^{-\sum_\alpha\beta_\alpha(
H_\alpha-\mu_\alpha N_\alpha)}$. After $t_0$ one can turn on the
tunneling couplings between the dot and the leads to let the system
evolve. In comparison with the partition-free scheme, we also shift
each energy level in lead $\alpha$ by $U_\alpha$ to preserve the
charge neutrality, i.e. $\epsilon_{\alpha k}\rightarrow
\epsilon_{\alpha k}+U_\alpha$, and let
$\beta_L=\beta_R=\beta$.\cite{Stef04195318} Then the non-local time
correlation function $g_\alpha(t,\tau)$ is the same as Eq.~(\ref{g})
with the same spectral density of Eq.~(\ref{gamma}). The spectral
Green function $u(\tau,t_0)$ has the same solution as the one in the
partition-free scheme. To demonstrate the initial-state differences
between the partition-free and partitioned schemes, we consider an
initial empty dot in the partitioned scheme. Thus, the initial
correlations for the partitioned scheme are
\begin{subequations}
\begin{align}
\langle a^\dag(t_0)a(t_0)\rangle&=0, \\
\langle a^\dag(t_0)c_{\alpha k}(t_0)&=0, \\
\langle c^\dag_{\alpha'k'}(t_0)c_{\alpha
k}(t_0)\rangle&=\delta_{\alpha k,\alpha'k'}
f_\alpha(\epsilon_{\alpha k}).
\end{align}
\end{subequations}
In this case, the non-local time system-lead correlation function
$g^\chi_\alpha(\tau,t_0)=0$, and the non-local time lead-lead
correlation function is given by,
\begin{align}
g^v_\alpha(t,\tau)&=\int\frac{d\epsilon}{2\pi}\Gamma_\alpha(\epsilon)f_\alpha(\epsilon+U_\alpha)e^{-i(\epsilon+U_\alpha)(\tau-t_0)},
\end{align}
which is indeed just the first term in
Eq.~(\ref{initialcorrelations}c), as an incoherent thermal effect
from lead $\alpha$.

\subsection{Transient transport dynamics}
The purpose of this subsection is to inspect how the localized bound
states and the applied bias affect the transient transport dynamics
for the partition-free and partitioned schemes. The transient
transport dynamics can be rationally described by the density matrix
(i.e. electron occupation in the single-level quantum dot) and the
transient transport current,
\begin{subequations}
\begin{align}
\rho^{(1)}(t)=& u(t,t_0)\rho^{(1)}(t_0)u^{*}(t,t_0)
+v'(t,t), \\
I_\alpha(t)=& -2e{\rm Re}
\big[\int_{t_0}^t d\tau [g_\alpha(t,\tau)\rho^{(1)}(\tau, t) \notag\\
&-\widetilde{g}'_\alpha(t,\tau)u^{*}(t,\tau)] +g^\chi_\alpha(t,t_0)
u^{*}(t,t_0) \big].
\end{align}
\end{subequations}
For the partition-free system, initially, the whole system is in
equilibrium with temperature $k_BT$, and chemical potential $\mu$.
The source and the drain of the partitioned system are prepared at
the same temperature ($k_BT_L=k_BT_R=k_BT$). The systems are said to
be unbiased under the condition $eV_{SD}=\mu_L-\mu_R=U_L-U_R=0$.

\begin{figure*}
\centerline{\scalebox{0.65}{\includegraphics{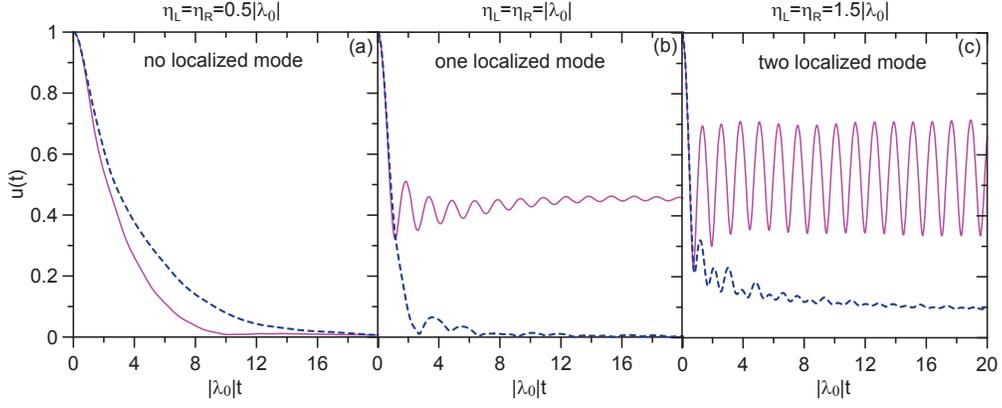}}} \caption{The
absolute value of time-dependent spectral Green function $|u(t)|$ at
different coupling ratios (a) $\eta_{L}=\eta_{R}=0.5|\lambda_0|$,
(b) $\eta_{L}=\eta_{R}=1.0|\lambda_0|$, (c)
$\eta_{L}=\eta_{R}=1.5|\lambda_0|$ with zero bias
$eV_{SD}=\mu_L-\mu_R=U_L-U_R=0$ (magenta solid line) and a finite
bias $eV_{SD}=\mu_L-\mu_R=U_L-U_R=3|\lambda_0|$ (navy blue dash
line). The energy level of the quantum dot
$\varepsilon_c=3|\lambda_0|$, and the band center of the two leads
$\epsilon_0=2.5|\lambda_0|$. For the unbiased case,
$\mu_L=\mu_R=2.5|\lambda_0|$, and $U_L=U_R=1.5|\lambda_0|$. For the
biased case, $\mu_L=4|\lambda_0|$, $\mu_R=|\lambda_0|$,
$U_L=3|\lambda_0|$, and $U_R=0$. We take $t_0=0$}\label{u}
\end{figure*}

The time evolution of the absolute value of the spectral Green
function $|u(t)|=|u(t,t_0=0)|$ which describes electron dissipation
in the dot system is the same for both the partition-free and
partitioned schemes in the unbiased and biased cases, as shown in
Fig.~\ref{u}.
For the unbiased case, the spectral Green function purely decays to
zero when there is no localized bound state, namely, the total
coupling ratio $\eta^2<2-\frac{\Delta}{|\lambda_0|}$. When one
localized bound state appears, i.e.
$2-\frac{\Delta}{|\lambda_0|}\leq \eta^2
<2+\frac{\Delta}{|\lambda_0|}$, $|u(t)|$ oscillates with time and
then approaches to a non-zero constant value in the steady-state
limit. This phenomenon becomes stronger when two localized bound
states occur for $\eta^2\geq2+\frac{\Delta}{|\lambda_0|}$. One can
find that $|u(t)|$ will oscillate in time forever when two localized
bound states occur simultaneously.

When the system is biased, $|u(t)|$ decays slower in comparison with
the unbiased system if the localized bound state has not appeared.
But when one localized bound state occurs, $|u(t)|$ decays much
faster and eventually approaches to zero. In other words, the effect
of the localized bound state can be manipulated by the bias voltage.
When two localized bound states appear, the situation is
dramatically changed, $|u(t)|$ not only decays much faster in
comparison with the unbiased case, but also the oscillation in the
unbiased case is washed out in the steady-state limit, and $|u(t)|$
approaches to a constant value. Numerically, we found that the
amplitude of one of the localized bound states will be dramatically
reduced when the bias voltage is applied. In other words, the
applied bias can suppress the effect of one of the localized bound
states.

\begin{figure*}
\centerline{\scalebox{0.65}{\includegraphics{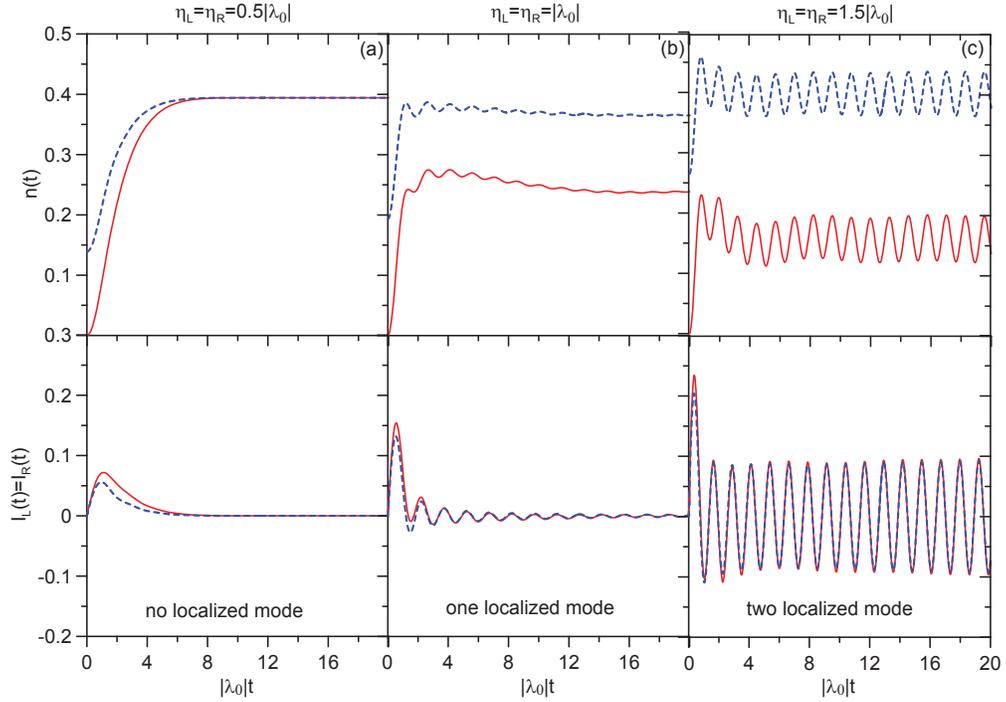}}}
\caption{The transient electron occupation of the dot and the
transient transport current for the unbiased case at different
coupling ratios (a) $\eta_{L}=\eta_{R}=0.5|\lambda_0|$, (b)
$\eta_{L}=\eta_{R}=|\lambda_0|$, (c)
$\eta_{L}=\eta_{R}=1.5|\lambda_0|$ for the partition-free (blue-dash
line) and partitioned (red solid line) schemes. The energy level of
the quantum dot $\varepsilon_c=3|\lambda_0|$, the band center of the
two leads $\epsilon_0=2.5|\lambda_0|$. For the partitioned scheme
the leads are prepared at $\mu_L=\mu_R=2.5|\lambda_0|$, and
$k_BT_L=k_BT_R=3|\lambda_0|$. For the partition-free scheme, the
system is initially at equilibrium with $\mu=|\lambda_0|$ and
$k_BT=3|\lambda_0|$. The applied bias voltage
$U_L=U_R=1.5|\lambda_0|$ after $t_0=0$.}\label{unbiased_case}
\end{figure*}

The above different dissipation dynamics in the biased and unbiased
cases will deeply change the transient and steady-state electron
occupation in the dot and also the transport current.
Fig.~\ref{unbiased_case} shows the electron occupation in the dot
and the transient transport current $I_L(t)=I_R(t)$ in the unbiased
case for the partitioned and partition-free schemes. The
partition-free system is initially at equilibrium so that the dot
contains a fraction of an electron, while the dot is initially empty
in the partitioned scheme. One can see that the effect of the
initial correlations vanish in the long-time limit when there is no
localized bound state. The steady-state electron occupation is the
same for the partition-free and partitioned schemes. However, when
one localized bound state appears, the electron occupation in the
dot oscillates in time for both the partitioned and partition-free
schemes, and approaches to different steady-state values. This
indicates that the initial correlations can affect on the density
matrix of the dot both in the transient and steady-state regimes due
to the existence of localized bound state. When two localized bound
states occur, it will generate an new oscillation with the frequency
being the energy difference of the two localized bound states for
the electron occupation and this oscillation will be maintained even
in the steady state,\cite{Stef07195115} where the
initial-correlation dependence becomes more significant, as shown in
Fig.~\ref{unbiased_case}(c) for the electron occupation. The
corresponding transient transport current for the partition-free and
partitioned schemes approach to the same value in a every short time
scale regardless whether the localized bound states exist or not.
The steady-state current approaches to zero if there is no localized
bound state. The current will also oscillate slightly when one
localized bound state occurs and approaches to zero in the steady
state. When both localized bound states appear, the current will
keep oscillations around zero forever.\cite{Stef07195115} However,
the initial-correlations dependence in the transport current is not
as significant as in the electron occupation in both the transient
and in the steady-state regimes, and even can be ignored in the
steady-state limit, as shown in Fig.~\ref{unbiased_case}.

\begin{figure*}
\centerline{\scalebox{0.65}{\includegraphics{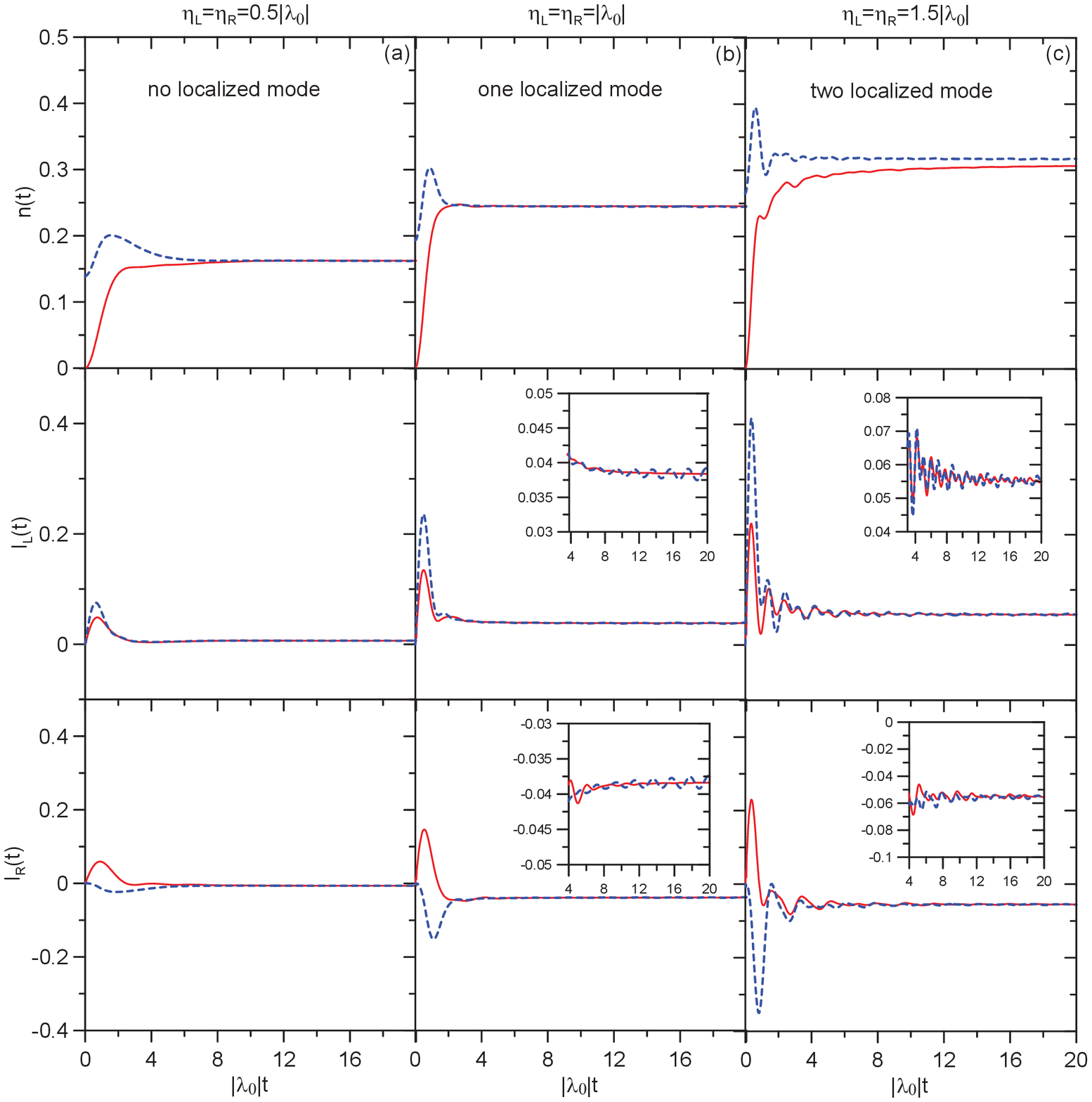}}}
\caption{The transient electron occupation of the dot and the
transient transport current for the biased case at different
coupling ratios (a) $\eta_{L}=\eta_{R}=0.5|\lambda_0|$, (b)
$\eta_{L}=\eta_{R}=|\lambda_0|$, (c)
$\eta_{L}=\eta_{R}=1.5|\lambda_0|$ for the partition-free (blue-dash
line) and partitioned (red solid line) schemes. The energy level of
the quantum dot $\varepsilon_c=3|\lambda_0|$, the band center of the
two leads $\epsilon_0=2.5|\lambda_0|$. For the partitioned scheme
the leads are prepared at $\mu_L=4|\lambda_0|$, $\mu_R=|\lambda_0|$,
and $k_BT_L=k_BT_R=3|\lambda_0|$. For the partition-free scheme, the
system is initially at equilibrium with $\mu=|\lambda_0|$ and
$k_BT=3|\lambda_0|$. The applied bias voltage $U_L=3|\lambda_0|$ and
$U_R=0$ after $t_0=0$.}\label{biased_case}
\end{figure*}

The time evolution of the electron occupation in the dot and the
transient transport current for both the partitioned and the
partition-free schemes for the biased case are shown in
Fig.~\ref{biased_case}. Compare Fig.~\ref{unbiased_case} with
Fig.~\ref{biased_case}, one can find that the applied bias restrains
most of the oscillation behavior in the electron occupation as well
as in the transport current, except for the very beginning of the
transient regime. Also, regardless of the existence of localized
bound states, the electron occupation and also the transport current
all approach to a steady-state value other than zero due to the
non-zero bias. In other words, the localized bound state has a less
effect on the electron occupation and the transport current when a
bias is applied. This is because, as we have pointed out in the
discussion of Fig.\ref{u}(c), the applied bias can reduce
significantly the amplitude of one of the localized bound states,
which suppress the oscillation of the transport electrons between
two localized bound states. However, the remaining localized bound
state will result in a small different steady-state values for
partition-free and partitioned schemes only in the electron
occupation. The corresponding transient current flow through the
left and right leads are quite different when a bias voltage is
applied. In particular, the transient transport current in the right
lead is positive in the beginning for the partitioned scheme because
the dot is initially empty, and it approaches to a negative
steady-state value in both schemes. But the steady-state current is
almost independent of the initial correlations as shown in the inset
graphs in Fig.~\ref{biased_case}. These result shows that the
initial correlation effects are not so significant in the
steady-state transport current in comparison with the electron
occupation.

\begin{figure*}
\centerline{\scalebox{0.5}{\includegraphics{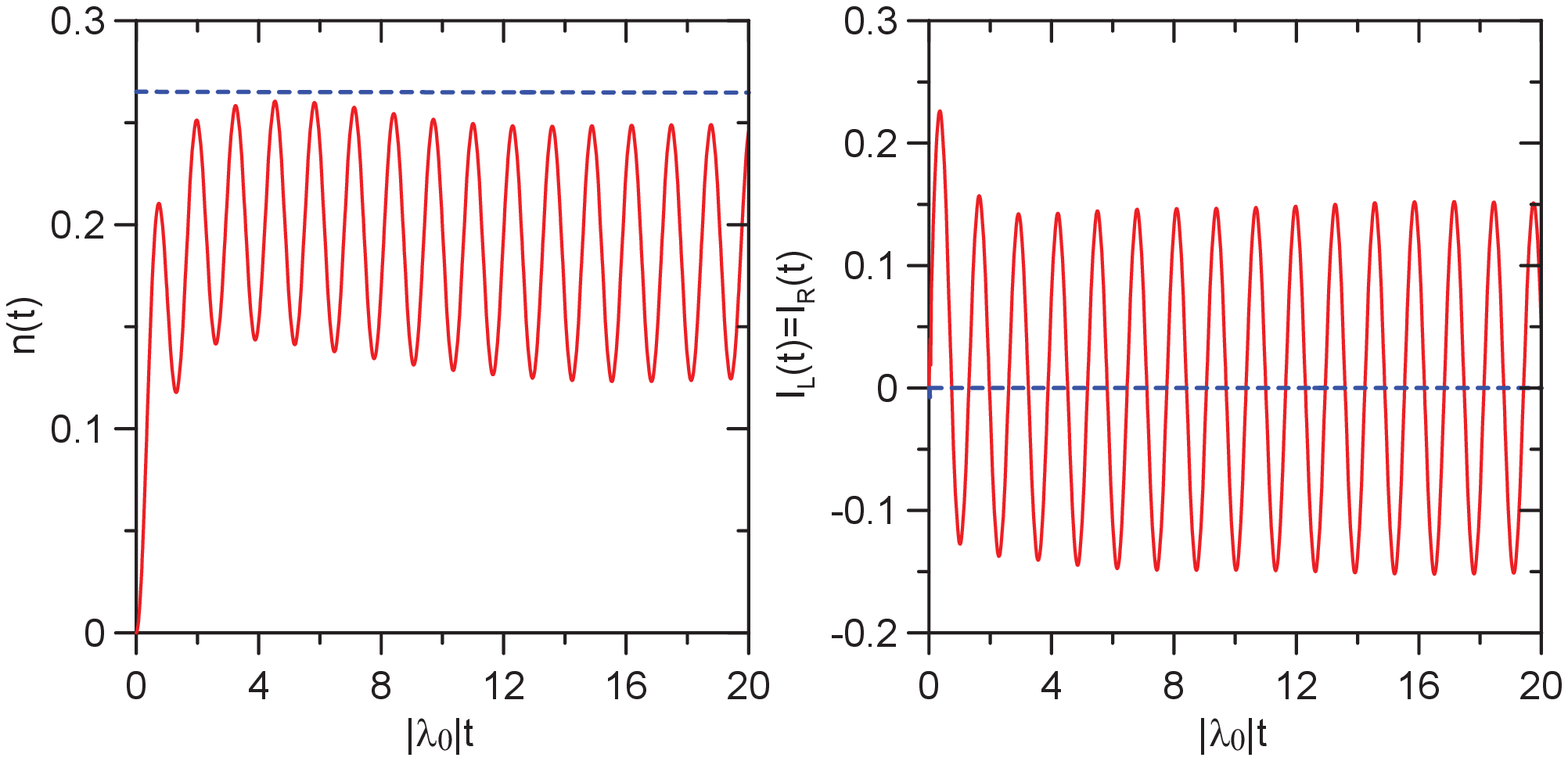}}} \caption{The
transient electron occupation $n(t)$ and the transient transport
current $I_{\alpha}(t)$ at the coupling ratio
$\eta_{L}=\eta_{R}=1.5|\lambda_0|$ with no energy shift
($U_L=U_R=0$) for the partition-free (blue dash line) and
partitioned system (red solid line). Other parameters are
$\mu=\mu_L=\mu_R=|\lambda_0|$ and
$k_BT=k_BT_L=k_BT_R=|\lambda_0|$}\label{U=0}
\end{figure*}

In the end, as a self-consistent check, we consider the case
$U_L=U_R=0$. For the partition-free scheme, the system should stay
in equilibrium. Our result is given in Fig.~\ref{U=0}, which is
expected. For the partitioned scheme, because of the appearance of
two localized bound states, the electron occupation in the dot will
keep oscillation around a steady-state value, and the system would
never reach to equilibrium with the leads. Also, the current
continuously oscillates with time around zero value. These results
agree with the results obtained in Ref.~[\onlinecite{Stef07195115}]
and Ref.~[\onlinecite{Dhar06085119}], in which one consider the
effect of localized bound state in quantum transport for
partition-free and partitioned schemes, respectively. In fact, these
results also agree with the fact that Anderson pointed out in
Anderson localization,\cite{Anderson} namely, the system cannot
approach to equilibrium when localized bound states occur.

\section{Summary and Discussion}
In summary, we investigate the transport dynamics of nanostructured
devices in the presence of initial system-lead and lead-lead
correlations. By deriving the exact master equation through the
extended quantum Langvein equation, the effect of initial
correlations is explicitly built into the fluctuation coefficients
in the master equation. The transient transport current
incorporating with initial correlations is deduced from the master
equation. Our transient transport theory based on master equation
approach derived in this paper is suitable for arbitrary initial
system-lead correlated state.

In application, we consider two initial states that are commonly
discussed in the literatures, one is the partition-free scheme in
which the initial correlations is presented explicitly. The other
one is the partitioned scheme in which there is no initial
correlations. The two schemes show the same steady-state behavior in
the small system-lead coupling regime. When the coupling between the
dot and the leads gets strong, localized bound states will appear,
and these two schemes give some different results both in the
transient regime and in the steady-state limit. It shows that the
initial correlation effects can not be ignored if the localized
bound states exist. Besides, when the localized bound states occur,
the device system can not approach to equilibrium with the leads,
and the initial correlation effects become more significant. These
initial correlation effects accompanied by the localized bound state
could be restrained by applying a finite bias voltage. Indeed, we
find that a finite bias can suppress the oscillation behavior in the
electron occupation and in the transport current induced by the
simultaneous existence of two localized bound states. Compare with
electron occupation in the dot, the steady-state current is
insensitive to the initial states, so the results are not so
distinguishable between the partitioned and the partition-free
schemes. Nevertheless, the transport currents are quite different in
the transient regime in these two schemes when the bias is applied.
Although we only consider a noninteracting nanostructure as a model
in the detailed exploration of the initial correlation effects, it
is not difficult to extend to include electron-electron interactions
in our theory because the coefficients of the master equation are
expressed in terms of the nonequilibrium Green functions which can
at least perturbatively or numerically handle the effect of
electron-electron interactions. We leave this problem for further
investigation.

\section*{Acknowledgment}
This research is supported by the National Science Conculs of ROC
under Contract No. NSC-102-2112-M-006-016-MY3, and the National
Center for Theoretical Science of Taiwan. It is also supported in
part by the Headquarters of University Advancement at the National
Cheng Kung University, which is sponsored by theMinistry of
Education of ROC.


\begin{references}

\bibitem{Haug2008} H. Haug and A. P. Jauho, \emph{Quantum Kinetics in Transport and Optics
of Semiconductors} (Springer Series in Solid-State Sciences
Vol.~123, 2008).

\bibitem{Wingreen1993} N. S. Wingreen, A. P. Jauho and Y. Meir, Phys. Rev. B
\textbf{48}, 8487(1993); \textit{idib.} \textbf{50}, 5528(1994).

\bibitem{Blanter2000} Y. M. Blanter, and M. B\"{u}ttiker, Phys. Rep \textbf{336}, 1-166 (2000).

\bibitem{Imry2002} Y. Imry, \emph{Introduction to Mesoscopic Physics}, (2nd Ed.
Oxford, 2002).

\bibitem{Datta1995}S. Datta, \emph{Electronic Transport in Mesoscopic
Systems}(Cambridge, 1995).

\bibitem{Schwinger1961} J. Schwinger, J. Math. Phys. \textbf{2}, 407 (1961);
L. V. Keldysh, Sov. Phys. JETP \textbf{20}, 1018 (1965).

\bibitem{Kadanoff1962}L.P. Kadanoff and G. Baym, \emph{Quantum Statistical
Mechanics}(Benjamin, New York, 1962)


\bibitem{Mahan1990} G. D. Mahan, \emph{Many Particle Physics} (Plenum, New
York, 1990), 2nd ed.

\bibitem{Buttiker1992} M. B\"{u}ttiker, Phys. Rev. B \textbf{46},
12485 (1992).

\bibitem{Meir1992} Y. Meir, and N. S. Wingreen, Phys. Rev. Lett. \textbf{68},
2512(1992).


\bibitem{Lu03422} W. Lu, Z. Ji, L. Pfeiffer, K. W. West, and A. J. Rimberg, Nature
(London) \textbf{423}, 422 (2003).

\bibitem{Bylander05361} J. Bylander, T. Duty, and P. Delsing, Nature (London) \textbf{434}, 361
(2005)

\bibitem{Gustavsson08152101} S. Gustavsson, I. Shorubalko, R. Leturcq, S. Sch\"{o}n, and K. Ensslin,
Appl. Phys. Lett. \textbf{92}, 152101 (2008).

\bibitem{Schoeller1994} H. Schoeller and G. Sch\"{o}n, Phys. Rev. B \textbf{50}, 18436 (1994).

\bibitem{Gurvitz1996} S. A. Gurvitz and Ya. S. Prager, Phys. Rev. B
\textbf{53}, 19532 (1996).

\bibitem{Li2005}X. Q. Li, J. Luo, Y. G. Yang, P. Cui, and Y. J. Yan, Phys. Rev. B \textbf{71},
205304 (2005).

\bibitem{Jin2008a} J. S. Jin, X. Zheng, and Y. J. Yan, J. Chem. Phys. \textbf{128}, 234703 (2008).

\bibitem{Tu2008} M. W. -Y. Tu and W. -M. Zhang, Phys. Rev. B \textbf{78}, 235311 (2008);
M. W. -Y. Tu, M.-T. Lee, and W. -M. Zhang, Quantum Inf. Processing
(Springer) \textbf{8}, 631 (2009).

\bibitem{Jin2010} J. S. Jin, M. W. -Y. Tu, W. -M. Zhang, and Y. J. Yan, New J. Phys. \textbf{12}, 083013 (2010).

\bibitem{Zhang12170402} W. M. Zhang, P. Y. Lo, H. N. Xiong, M. W. Y.
Tu, and F. Nori, Phys. Rev. Lett. \textbf{109}, 170402 (2012).

\bibitem{Tu2012} M. W.-Y. Tu, W.-M. Zhang, J. S. Jin, O. Entin-Wohlman, and A.
Aharony, Phys. Rev. B \textbf{86}, 115453 (2012); M. W.-Y. Tu, A.
Aharony, W.-M. Zhang, and O. Entin-Wohlman, Phys. Rev. B
\textbf{90}, 165422 (2014).

\bibitem{Moskalets2004} M. Moskalets and M. B\"{u}ttiker, Phys. Rev. B \textbf{69}, 205316 (2004).




\bibitem{Cini805887} M. Cini, Phys. Rev. B \textbf{22}, 5887 (1980).

\bibitem{Dhar06085119} A. Dhar and D. Sen, Phys. Rev. B \textbf{73},
085119 (2006).

\bibitem{Stef04195318} G. Stefanucci and C.-O. Almbladh, Phys. Rev.
B \textbf{69}, 195318 (2004).

\bibitem{Stef07195115} G. Stefanucci, Phys. Rev. B \textbf{75},
195115 (2007).


\bibitem{boundstate1} J. Taylor, H. Guo, and J. Wang, Phys. Rev. B
\textbf{63}, 245407 (2001).

\bibitem{boundstate2} P. Pomorski, L. Pastewka, C. Roland, H. Guo,
and J. Wang, Phys. Rev. B \textbf{69}, 115418 (2004).

\bibitem{boundstate3} V. Vettchinkina, A. Kartsev, D. Karlsson, and C.
Verdozzi, Phys. Rev. B \textbf{87}, 115117 (2013).

\bibitem{Anderson} P. W. Anderson, Phys. Rev. \textbf{109}, 1492
(1958).

\bibitem{FeynmanVernon} R. P. Feynman and F. L. Vernon, Ann. Phys. \textbf{24}, 118 (1963).

\bibitem{Yang14115411} P. Y. Yang, C. Y. Lin, and W. M. Zhang, Phys.
Rev. B \textbf{89}, 115411 (2014).


\bibitem{Schi9814978} A. Schiller and S. Hershfield, Phys, Rev. B
\text{58}, 14978 (1998).




\end{references}
\end{document}